# Frequently Asked Questions for: *The Atoms of Neural Computation*

Gary Marcus (NYU), Adam Marblestone (MIT), Tom Dean (Google)

One common view holds that that the human neocortex consists of a single, massively-repeated, "canonical" computation (Creutzfeldt, 1977; George & Hawkins, 2009; Hawkins & Blakeslee, 2007; Kurzweil, 2012; Mountcastle, 1978). But, after forty years, there is still no consensus about the nature of such a putative canonical computation, and little evidence that such uniform architectures can capture the diversity of cortical function in simple mammals, let alone language and abstract thinking (Hadley, 2009; Marcus, 2001). As we suggest in (Marcus, Marblestone, & Dean, 2014), an alternative is that the cortex might consist of *a diverse set of computationally distinct building blocks that implement a broad range of elementary, reusable computations*.

Sensory cortex, for example, might be rich in circuits instantiating primitives for hierarchical pattern recognition, like receptive fields, pooling and normalization; prefrontal cortex (PFC) might contain circuits supporting variable binding, sequencing, gating and working memory storage; motor cortex might specialize in circuits devoted to sequencing and coordinate transformation. Differences between building blocks might stem from fine-grained, area-specific variations in connectivity and plasticity rules, even in the context of a *superficially* stereotyped cortical anatomy as defined by the typical morphologies and laminar placements of neurons.

**What might these diverse computational "building blocks" actually do?**

We do not yet have a clear answer, and can provide only a preliminary "guess" at this point—efforts to develop an improved "taxonomy" of cortical computation are clearly needed.

Drawing on a wide survey of articles in cognitive and computational neuroscience, we present **Table 1** as a *preliminary, exploratory example* of what such a taxonomy might look like. For certain computations, such as variable binding (see below) or routing, there are many proposed neural implementations, and so multiple possibilities are given, though we have made no attempt to be exhaustive. Our commitment is not to the specific, but to the broader strategy, namely cataloguing a heterogeneous set of building blocks that could connect bottom-up neurophysiology and neural circuit analysis with candidate computations derived from top-down analysis (e.g., high-level cognitive modeling (Anderson, 2007)).

| Computation | Potential algorithmic/ representational realization(s) | Potential neural implementation(s) | Putative brain location(s) |
|---|---|---|---|
| Rapid perceptual classification | Receptive fields, pooling and local contrast normalization | Hierarchies of simple and complex cells (Anselmi et al., 2013) | Visual system |
| Complex spatiotemporal pattern recognition | Bayesian belief propagation (Lee & Mumford, 2003) | Feedforward and feedback pathways in cortical hierarchy (George & Hawkins, 2009) | Sensory hierarchies |
| Learning efficient coding of inputs | Sparse coding | Thresholding and local competition (Rozell, Johnson, Baraniuk, & Olshausen, 2008) or other mechanisms | Sensory and other systems |
| Working memory | Continuous or discrete attractor states in networks | Persistent activity in recurrent networks (X.-J. Wang, 2012) | Prefrontal cortex |
| Decision making | Reinforcement learning of action-selection policies in PFC/BG system | Reward-modulated plasticity in recurrent cortical networks coupled with winner-take-all action selection in the basal ganglia (Rajesh P N Rao, 2010) | Prefrontal cortex and basal ganglia |
| | Winner-take-all networks | Recurrent networks coupled via lateral inhibition (X.-J. Wang, 2012) | Prefrontal cortex |
| Routing of information flow | Context-dependent tuning of activity in recurrent network dynamics | Recurrent networks implementing line attractors and selection vectors (Mante, Sussillo, Shenoy, & Newsome, 2013) | Common across many cortical areas |
| | Shifter circuits | Divergent excitatory relays and input-selective shunting inhibition in dendrites (Olshausen, Anderson, & Van Essen, 1993) | Common across many cortical areas |
| | Oscillatory coupling (Colgin et al., 2009; Ketz, Jensen, & O'Reilly, 2014) and phase synchronization (Salinas & Sejnowski, 2001) | Frequency filtering via feedforward inhibition (Akam & Kullmann, 2010) or selective effects on spike coherence in excitatory/inhibitory networks with multiple classes of interneurons (Börgers, Epstein, & Kopell, 2008) | Common across many cortical areas |
| | Modulating excitation/inhibition balance during signal propagation | Selective modulation of inhibitory neurons in balanced networks, e.g., via cholinergic modulation (Vogels & Abbott, 2009) | Common across many cortical areas |
| Gain control | Divisive normalization | Shunting inhibition in networks, balanced background synaptic excitation and inhibition or other mechanisms (Carandini & Heeger, 2012) | Common across many cortical areas |
| Sequencing of events over time | Feed-forward cascades | Synfire chains (Ikegaya et al., 2004) | Language and motor areas |
| | Serial working memories | Ordinal serial encoding through variable binding (Choo and Eliasmith, 2010) | Prefrontal cortex |
| Representation and transformation of variables | Population coding (Georgopoulos, Schwartz, & Kettner, 1986), probabilistic population coding (Ma, Beck, Latham, & Pouget, 2006), or many variants | Time-varying firing rates of neurons representing dot products with basis vectors, nonlinear (Shamir & Sompolinsky, 2004) or probabilistic variants (Ma et al., 2006), and generalizations to high-dimensional vectors (Eliasmith et al., 2012) | Motor cortex and higher cortical areas |
| Variable binding | Indirection (Kriete, Noelle, Cohen, & O'Reilly, 2013) | Patterns of cortical neural activity in an area encode "pointers", and gate the activity of other pointer-specified areas via projections to and from the basal ganglia | Prefrontal cortex / basal ganglia loops |
| | Dynamically partionionable autoassociative networks (Hayworth, 2012) | Selective gating of subsets of large-scale attractor networks linking many cortical regions | Higher cortical areas |
| | Holographic reduced representations (Eliasmith et al., 2012) | Circular convolution applied to population coded vector representations in cortical working memory buffers, gated by thalamic signals | Higher cortical areas |

**Table 1:** Preliminary, illustrative example of a proposal for a partial taxonomy of elementary, reusable operations (left column) that could underlie the diversity of cortical computation, with possible neural implementations (middle two columns) and associated brain structures (right column), based on a survey of mechanisms from the existing computational neuroscience literature. Key sources influencing the overall breakdown include (Eliasmith, 2013; O'Reilly, 2006; Rajesh P.N. Rao, Olshausen, & Lewicki, 2002). The differential configuration of such circuit primitives in a given cortical area could be shaped by wiring rules that are both activity dependent and cell type specific, with variation from area to area driven by molecular cues and differential cell type compositions as well as by the statistics of the "input".

**The cortex is just a part of the picture, right?**

Yes! Connections to other brain areas like the basal ganglia (BG) (Gurney, Prescott, & Redgrave, 2001) and thalamus (Sherman, 2007) likely contribute (Buschman & Miller, 2014) to sequential and recursive processing capabilities that require state-dependent routing of information between cortical areas. As another example, the cortex is closely coupled to the hippocampus (Hasselmo & Wyble, 1997; Treves & Rolls, 1992) to enable various forms of memory storage and retrieval, which may also be coordinated via the thalamus (Ketz et al., 2014). The interplay of these areas may be crucial to many cognitive processes.

**What is "variable binding" and how might it be implemented in the brain?**

There are several qualitatively different models for how neural circuitry might implement *variable binding* -- the transitory or permanent tying together of two bits of information: a variable (such as an *X* or *Y* in algebra, or a placeholder like *subject* or *verb* in a sentence) and an arbitrary instantiation of that variable (say, a single number, symbol, vector, or word).

*PFC/BG indirection:* One hypothesis is that the basal ganglia can control or "gate" the storage and processing of information in the PFC. The crucial feature of the PFC/BG indirection model for variable binding (Kriete et al., 2013) is that, in addition, activity patterns at a particular location in the PFC can instruct the basal ganglia as to *which other locations* in the PFC should be gated. Thus, activity patterns in PFC area A can encode "pointers" or "addresses" that direct operations on PFC area B -- such as storage, updating or outputting -- irrespective of the *contents* of PFC area B.

*Dynamically partitionable autoassociative neural networks (DPANN):* So-called "anatomical binding" models propose that each variable corresponds to an anatomically defined region or register (see also (Marcus, 2001)). The major questions for such models pertain to how "global" symbols are defined. For instance, the same symbol "black" should apply equally to the concepts "black cat" and "black sedan", even though cats (animals) and sedans (cars) may be represented in widely separated regions of the cortex. The Dynamically Partionable Auto-Associative Neural Network (DPANN) model (Hayworth, 2012) proposes that such global symbols arise from a large-scale network linking many regions of the brain. A given symbol then corresponds to a global attractor state of the dynamics of this integrated network. To perform syntactic tasks, subsets of the synapses in this network are turned on and off ("gated"), leading to the organized transmission of information between particular sets of registers.

This model is designed to subserve the production rule matching operations assumed in higher-level cognitive modeling frameworks like ACT-R (Anderson, 2007), which require the ability to evaluate an arbitrary number of variablized rules in a single 50 millisecond production cycle; implementing ACT-R arguably also requires true role-filler independence of symbols, which would be enabled by a DPANN binding system (Ken Hayworth, personal communication).

An appealing feature of this model is that it relies only on "biologically plausible" neural mechanisms such as Hebbian learning, gating of information flow, and attentional "spotlights". At the same time, it predicts the existence of "equality detection neurons", "feature detection neurons" and "transfer control neurons" as well as the DPANN attractor network itself. Generalizations of this idea to different types of attractor networks may be of interest.

*Vector symbolic architectures:* In (Eliasmith et al., 2012), the collective activity pattern of a group of neurons encodes a high-dimensional vector. These vectors are transformed in mathematically precise ways by the synaptic connectivity between different neural populations. By representing variables as vectors, like **subject**, **object** or **verb**, and also representing potential role fillers as vectors, like **dog** or **car**, the synaptic connectivity between groups of neurons can be chosen so as to implement the "compositional" processes characteristic of symbol manipulation. For instance, assigning "dog" to the role of "subject" in a sentence might occur by multiplying the two vectors **subject** and **dog**, to form the activity pattern **subject** $\otimes$ **dog**. Importantly, if the corresponding groups of neurons were instead activated with activity patterns representing **subject** and **cat**, the exact same synaptic connectivity between populations would generate the composition **subject** $\otimes$ **cat**. It is not yet clear whether the precise synaptic connectivity patterns required by such "vector symbolic architectures" actually exist in the brain.

*Binding through synchrony:* Variables (roles) represented in one brain area are linked to values (fillers) in other brain areas via the synchronous oscillation of both areas (Lokendra Shastri, 1993). Multiple roles can be represented in a role area, and multiple fillers in a filler area, so long as multiple oscillations can co-exist that are mutually out of phase, allowing a role to be mapped to a filler via phase synchrony.

Experimental designs (Dong, Mihalas, Qiu, von der Heydt, & Niebur, 2008) are sought that could distinguish among these qualitatively different neural substrates of variable binding, or others.

**Don't experiments already prove that the cortex is uniform?**

One of the most cited sources of evidence in favor of putative cortical uniformity comes from a classic series of experiments by Sur and collaborators (Sharma, Angelucci, & Sur, 2000), based on (Frost & Metin, 1985), in which visual inputs to primary visual cortex (V1) were rerouted to the primary auditory cortex (A1), which in turn was shown to be capable of processing visual stimuli. While these studies are often taken to imply a "uniform" cortical substrate, several caveats are in order (Marcus, 2008). First, such results have only been demonstrated within primary sensory cortices, which might plausibly share a somewhat uniform pattern recognition architecture, whereas other areas (e.g., in frontal cortex) may be highly diverged from such pattern recognizers, as we emphasize here. Second, the "rewired" auditory cortex still retains some of its intrinsic properties (Majewska, Newton, & Sur, 2006) and the resulting "visual" system is not without defects. Visual input leads only to a partial re-structuring of A1 at an anatomical level (Sharma et al., 2000). Third, the areas were not (contrary to a widespread characterization) directly "rewired"; rather intrinsic axon guidance mechanisms used in development were harnessed to guide rewiring, in part relying on molecular cues shared between visual and auditory areas, but which might not be equally effective if visual input were induced to connect to, e.g., prefrontal cortex. In the subsequent decade, there appears not to have been any published report of successful attempts to reroute visual inputs to other areas that seem more different (e.g., prefrontal cortex), indirect evidence that cortical "interchangeability" may be far from general.

**How could biology make a heterogeneous cortex, rather than repeating a single module?**

The emergence of a taxonomy of structurally and functionally heterogeneous computational building blocks fits with biology's tendency to accrete complexity during evolution through processes such as duplication and divergence (Belgard & Geschwind, 2013). Central pattern generators, for instance, may have diversified over evolutionary time (Yuste, MacLean, Smith, & Lansner, 2005), thereby giving rise to a broad variety of related but distinct computational primitives. Biology includes a rich set of mechanisms through which a heterogenous array of distinct computational circuitry, varying between cortical areas, could be programmed, even in advance of experience, and tuned through a mixture of molecular cues and activity-dependent plasticity rules.

Genetically guided processes of developmental biology, such as cell division, cell migration and axon guidance could differentially shape local circuitry. Such processes can be extremely precise. There are a vast range of potential such processes, and surely more remain to be discovered. For example, families of molecules exhibiting combinatorial molecular recognition, such as cadherins (Duan, Krishnaswamy, De la Huerta, & Sanes, 2014) or the many RNA splice isoforms of Dscam (Hattori et al., 2007) in the fly, may endow neurons with locally unique "identity tags" to constrain their connectivity. Conjunctions of axon-guidance gradients and competition between axons can also yield highly precise structures, such as topographic maps (Simpson & Goodhill, 2011; Triplett et al., 2011). Genes such as SAM68 (Iijima et al., 2011) can govern the activity-dependent alternative splicing of neurexin molecules (Treutlein, Gokce, Quake, & Südhof, 2014) that play important functions in the formation, maturation, and maintenance of synapses. The state of pre-synaptic neurexin splicing can even influence post-synaptic receptor trafficking (Aoto, Martinelli, Malenka, Tabuchi, & Südhof, 2013), potentially allowing for a configurational code that could integrate intrinsic and extrinsic cues; such neurexins are represented differentially across the brain, and across developmental stages (Iijima et al., 2011). Neurexins also interact with neuroexophilin ligands, which are differentially expressed in sub-populations of synapses (Born et al., 2014), to influence synaptic function.

At a theoretical level, biologically plausible developmental rules could also generate certain types of precise connectivity, such as perfectly reciprocal connections between areas, by "sharpening" connections (Cook, Matthew and Jug, Florian and Krautz, 2011). Combinations of spike-timing-dependent Hebbian plasticity and heterosynaptic competition could contribute to the sculpting of connectivity underlying precise sequencing of neural activity (Fiete, Senn, Wang, & Hahnloser, 2010). There is a vast range of other possible mechanisms for differential formation of precise circuits in different areas.

One possibility is that statistically stereotyped geometric templates (e.g., cortical columns), that are defined by canonical neuron morphologies and their laminar placements (Hill, Wang, Riachi, Schürmann, & Markram, 2012), may be largely invariant across the cortex, but with specific local synaptic connectivity fine-tuned by a mixture of plasticity and area-specific molecularly guided developmental rules. Such rules, by hypothesis, would guide the "last mile" of neural wiring in different blocks, by, for example, directing the local formation, removal or change in size of synapses "on top" of otherwise stereotyped arborization. For instance, layer 5 pyramidal cells with molecular tag α might seek a specific molecular cue α' in a transiently connected cell to drive synapse growth or pruning, whereas layer 5 pyramidal cells with molecular tag β might adjust synaptic weights with their neighbors via Hebbian plasticity mechanisms. This molecularly defined heterogeneity could allow *different circuit configurations to emerge from circuitry that superficially appears, without knowledge of the detailed local connectivity, to be stereotyped*. Molecular properties of cells with the same morphological cell type can differ even within a single cortical layer (Khazen, Hill, Schürmann, & Markram, 2012), making it possible in principle that local synaptic connectivity might be mediated not just by activity-driven plasticity (Yuste et al., 2005) but also by different molecularly defined cell sub-types of the traditionally recognized, chemo-morphologically defined cell types. Relatedly, there is evidence that some neuronal morphologies and placements evolved to be consistent with many potential patterns of synaptic connectivity (Kalisman, Silberberg, & Markram, 2005).

**Is there any biological evidence that would support the possibility of cortical diversity?**

At a coarse level, there are several well-known cytoarchitectonic differences such as the agranularity of motor cortex layer 4, and a rostral-caudal gradient in supra-granular (II-IV) layer neuron numbers per unit of cortical surface115. Likewise the distribution of different types of interneurons can differ sharply between areas: in V1, PV-containing interneurons (including fast-spiking basket cells) are prevalent (~75%) relative to (CB- and CR-containing) interneurons, whereas in the prefrontal cortex CR-containing neurons (about 45%) outnumber both PV and CB-containing interneuron's (X.-J. Wang, 2012). Likewise, canonical structural features such as ocular dominance columns appear in some parts of cortex but not in most others, and are present in some closely-related species but not in others (Horton & Adams, 2005).

Local micro-circuitry also differs between cortical areas. For example, motor cortex is dominated by "top-down" layer 2 to layer 5 local connections (Weiler, Wood, Yu, Solla, & Shepherd, 2008), while primary sensory areas have prominent "ascending" local connections L4 to L2/L3 and L5 to L2 (Shepherd & Svoboda, 2005). Synaptic connectivity and synaptic properties differ between frontal cortex and primary visual cortex. The probability of recurrent connections between neurons in frontal cortex is substantially higher than in the primary sensory cortex (X.-J. Wang, 2012). Pyramidal cells in prefrontal cortex have, on average, up to 23 times more dendritic spines than those in the primary visual areas (Elston, 2003) and cells in visual association areas are larger and have more spines than in primary visual areas (Elston, 2002). Whereas most excitatory cortical synapses exhibit short-term synaptic depression, some excitatory synapses in the frontal cortex exhibit short-term facilitation, perhaps contributing to the generation of the sustained activity characteristic of prefrontal processing (Y. Wang et al., 2006). Recent work suggests there can also be important microcircuit-level differences within an area, such as a 2.5 fold difference in the number of neurons per cortical barrel column from dorsal to ventral in rodent S1 (Meyer et al., 2013).

There are differences between gene expression in different areas of the cortex such that "the spatial topography of the neocortex is strongly reflected in its molecular topography—the closer two cortical regions, the more similar their transcriptomes" (Hawrylycz et al., 2012). Data from the Allen Mouse Atlas shows that specific genetic markers, such as RAR-related orphan receptor beta, potassium voltage-gated channel, subfamily H (eag-related) member 7, ephrin A5, and activity regulated cytoskeletal-associated protein (Arc) are expressed at markedly higher levels in primary sensory areas (V1, S1, and A1) than elsewhere in the cortex (Stefan Mihalas, personal communication). Stansberg et al (Stansberg, Ersland, van der Valk, & Steen, 2011) review 65 genes that are enriched in specific cortical regions.

There are functional differences between cortical areas that may not be attributable to activity alone; neural activity in frontal areas, for example, tend to be less immediately stimulus-driven and more persistent than primary sensory areas (X.-J. Wang, 2012); at the same time, primary sensory areas differ significantly from one another in their sensitivity to brief temporal offsets (Yang & Zador, 2012).

Recent investigations of mouse primary somatosensory cortex (Sorensen et al., 2013) combining in situ hybridization data mining, marker gene co-localization, and retrograde tracing suggest that particular subsets of cell types, such as subsets of layer 5 pyramidal cells, project differently to distal targets – in a fashion that is systematically governed by a molecular, combinatorial code. Whether such a combinatorial code applies at the local microcircuit level remains unknown.

Collectively, this evidence makes it plausible that different areas of cortex might be wired in different ways, despite a shared six-layered substrate, potentially supporting qualitatively distinct computations.

**What about the human cortex vs. that of other mammals: isn't that just a matter of size?**

Even if most mammals manage with a limited form of cortical computation based only on motifs like pattern recognition and working memory, humans might further re-purpose elements such as the "cortical column", potentially providing for the capacity to concatenate "arbitrary" symbols (Marcus, 2001) or represent recursive structures (Hauser, Chomsky, & Fitch, 2002; Traxler, Boudewyn, & Loudermilk, 2012). Seemingly slight molecular differences, such as the alternative splicing of GPR56, may have profound consequences in human development and evolution, for example by selectively altering stem cell proliferation and gyral patterning in the vicinity of Broca's area (Bae et al., 2014). Intriguingly, recent transcriptome analyses (Hawrylycz et al., 2012; Konopka et al., 2012) show that the human frontal areas are marked by "a predominance of genes differentially expressed within human frontal lobe and a striking increase in transcriptional complexity specific to the human lineage in the frontal lobe" – exactly as one might expect if some new computational configuration had recently evolved.

**What is the status of "canonical microcircuit" models of the cortex?**

While we will not attempt a comprehensive survey here, several theories (Anselmi et al., 2013; Bhand, Mudur, Suresh, Saxe, & Ng, 2011; George & Hawkins, 2009; Mallat, 2013; Poggio, Mutch, Leibo, Rosasco, & Tacchetti, 2012) aim to give a general account of a broad range of hierarchical sensory computations. For example Anselmi et al (Anselmi et al., 2013) offer a principled account of how rapid computation of transformation-invariant signatures of new inputs by the ventral visual stream could underlie recognition or category learning from small numbers of examples; it is unclear to us, however, whether these or similar mathematical principles would generalize beyond the sensory hierarchies, to other areas of the cortex (e.g., prefrontal) and aspects of cognition (e.g., language).

Maass (2000) studied the computational versatility of soft winner-take-all or soft-MAX operations, which were suggested by Douglas and Martin to be central to the operation of the putative canonical cortical microcircuit (Douglas & Martin, 2004) and which also feature heavily in a variety of other models. Recent work suggests that spike-timing dependent plasticity on top of a soft-MAX circuit structure can give rise to computations approximating simple Bayesian inference (Kappel, Nessler, & Maass, 2014) as well as hidden Markov models (Kappel et al., 2014). Kouh and Poggio (Kouh & Poggio, 2008) have also suggested tunable circuits that approximate MAX-like operations for some parameter values but generalize to other nonlinear operations under different parameters.

"Liquid state machines" (Maass, Natschläger, & Markram, 2002) are tunable computational elements that are consistent with random internal wiring, and it has been suggested that this framework may provide a canonical, but highly configurable, cortical microcircuit model. The basic idea is to exploit a randomly wired recurrent network of nonlinear elements, which will generate a rich internal repertoire of nonlinear functions. A subset of these elements is chosen and their outputs run through tunable linear decoders. Merely by tuning the weights of the linear decoders, one can synthesize a wide range of computations on spatiotemporal input streams. These networks can be trained to perform multiple distinct functions simultaneously using Hebbian plasticity modulated by a simple global reward signal (Hoerzer, Legenstein, & Maass, 2014), suggesting the possibility that some computational logic blocks could be configured through reinforcement learning (though others may be prewired through the use of molecular cues).

**Don't there already exist cortical models incorporating a range of computing primitives?**

To some extent, but we'd like to see more; our hope is that more researchers will move towards developing a richer, more fully articulated, better empirically and theoretically justified version of **Table 1**.

An incomplete list of steps that might be in the right direction include the following:

- The semantic pointer architecture unified network (SPAUN) system (Eliasmith et al., 2012) is a single spiking neural network model which generates diverse functions and flexibly integrates these functions in a manner suggestive of the flexibility of primate cognition. SPAUN combines serial working memory (via a recurrent attractor neural network with a family of fixed points corresponding to stored values (Singh & Eliasmith, 2006), reinforcement learning, action selection (Stewart, Bekolay, & Eliasmith, 2012), symbol manipulation via vector symbolic architectures (Terrence C. Stewart, Xuan Choo, 2010), motor control (DeWolf & Eliasmith, 2011), image recognition and various forms of pattern completion, and arguably comes closest, among existing models, towards attempting to provide an integrated mesoscale theory of cortical organization. SPAUN models visual cortex as a hierarchy of feature detectors (restricted Boltzmann machines), and prefrontal cortex as a kind of working memory. All these systems are implemented in spiking neurons through a common mathematical framework, the Neural Engineering Framework (NEF) (Eliasmith & Anderson, 2004), for representation and transformation of variables in neuronal populations.
- The deep reinforcement learning algorithm of (Mnih et al., 2013) combines Q-learning methods from the field of reinforcement learning with deep convolutional networks, as well as a hippocampus-inspired "experience replay" mechanism (which "smoothes out" the learning process as compared to standard sequential Q-learning), in order to learn mappings from "sensory" data to optimal actions (mapping the last few screen images to the predicted reward values of a fixed set of actions).
- The LEABRA model divides cortex into posterior sensory hierarchies and frontal areas which perform "active maintenance" of working memory representations used in decision making, while also incorporating a basal ganglia model that gates information flow in the prefrontal cortex and learns via reinforcement (Petrov, Jilk, & O'Reilly, 2010).
- The Neural Turing Machine (NTM) (Graves, Wayne, & Danihelka, 2014) offers a novel approach to integrating location-addressable memory functions inspired by biological working memory, attention and variable binding with recurrent neural networks that can be trained by gradient descent. This work also highlights the relationships between long short-term memory (LSTM) networks (Hochreiter & Schmidhuber, 1997) and models of PFC/basal ganglia functions (Hazy, Frank, & O'Reilly, 2006; Krueger & Dayan, 2009) that include gating and persistent working memory. When an LSTM is trained to control a memory that is jointly addressable by content and by location (using a mechanism inspired by the notion of "attention), the resulting networks seem to learn sequential algorithms that require variable binding and indirection (Graves et al., 2014). The recurrent loops carried out by LSTM and NTM networks are suggestive of ACT-R production rules (Anderson, 2007) and hypothesized cognitive cycles in the brain (Buschman & Miller, 2010). See also (Bahdanau, Cho, & Bengio, 2014; Weston, Chopra, & Bordes, 2014).

But none of these go far enough. Many are still framed as one-size-fits-all solutions; our goal is not to start from a single algorithmic task (e.g., game playing, as in the deep reinforcement learning algorithm), nor to propose a single master algorithm for deriving networks by gradient descent optimization (as in the NTM) or via a single mathematical framework (as in the NEF), nor with a single canonical operation or learning rule, but instead to help reframe the research enterprise – as one in which a core goal is to develop a taxonomy of mechanisms, akin to the basic instructions in a microprocessor, i.e., to encourage cross-disciplinary efforts towards characterizing an integrative taxonomy and phylogeny of neural computations in the brain: what such circuits achieve, how they are realized biologically and computationally, and how they have diverged over evolutionary time.